\documentclass[authoryear,preprint,review,12pt]{elsarticle}

\usepackage{amssymb}
\usepackage{graphicx}
\usepackage{color}
\newcommand{\logg}{\ensuremath{\log g}}                                 
\newcommand{\kms}{\,km\,s$^{-1}$}                                       

\journal{New Astronomy}

\begin{document}

\begin{frontmatter}



\title{Analysis of the Massive Eclipsing Binary V1441\,Aql}

\author[l1]{\"{O}. \c{C}ak{\i}rl{\i} \corref{cer}}

\address[l1]{Ege University, Science Faculty, Astronomy and Space Sciences Dept., 35100 Bornova, \.{I}zmir, Turkey \corref{cer}}
\cortext[cer]{Corresponding author. Tel.: +902323111740; Fax: +902323731403 \\
E-mail address: omur.cakirli@gmail.com}

\author[l1]{C. Ibanoglu}
\author[l1]{E. Sipahi}
\author[l3]{A. Frasca}
\author[l3]{G. Catanzaro}
\address[l3]{Osservatorio Astrofisico di Catania, Via S. Sofia 78, 95123 Catania, Italy}

\begin{abstract}
We present new spectroscopic observations of the early type, double-lined eclipsing binary 
V1441\,Aql. The radial velocities and the available photometric data obtained by 
$ASAS$ is analysed for deriving the parameters of the components. The components of 
V1441\,Aql are shown to be a B3\,IV primary with a mass M$_p$=8.02$\pm$0.51 M$_{\odot}$ and radius 
R$_p$=7.33$\pm$0.19 R$_{\odot}$ and a B9 III secondary with a mass M$_s$=1.92$\pm$0.14 M$_{\odot}$ 
and radius R$_s$=4.22$\pm$0.11 R$_{\odot}$. Our analyses show that V1441\,Aql is a double-contact 
system with rapidly rotating components. Based on the position of the components plotted on the 
theoretical Hertzsprung-Russell diagram, we estimate that the ages of V1441\,Aql is about 
30\,Myr, neglecting the effects of mass exchange between the components. Using the UBVJHK magnitudes
and interstellar absorption we estimated the mean distance to the system V1441\,Aql as 550$\pm$25\,pc. 
\end{abstract}


\end{frontmatter}

\section{Introduction}\label{sec:intro}
High-mass stars are much less frequent than intermediate- or low-mass stars due to both the star formation
process, which gives rise to an initial mass function  declining with the mass \citep[e.g.][]{Salpeter1955,Kroupa2001},
and to their shorter evolutionary times. However, high-mass stars are very important because they can affect their 
surroundings with their winds, their strong radiation fields, and their catastrophic death as supernovae, chemically 
enriching their environment and triggering star formation. They usually form within the dense cores of stellar 
clusters and/or associations where dynamical interactions play an important role. It is widely believed that 
massive stars may be the product of collisions between two or more intermediate-mass stars. This idea is supported 
by the fact that a large fraction of massive stars harbour close companions, as failed mergers. It has been recently 
estimated that at least 50\,\% of massive stars are member of binary or multiple star system \citep{San12}. This lucky 
occurrence allows to directly measure the masses by means of their radial velocity (RV) curves. In many cases, spectral 
lines of both components are visible (SB2 systems), allowing to derive the orbital parameters like the period, 
$P_{\rm orb}$, the projected semi-major axes, $a_{1,2}\sin i$, and the masses, $M_{1,2}\sin^3i$, apart from the 
factor $\sin^3i$. If our line-of-sight is close to the orbital plane and fractional radii of the components are 
not too small the stars display mutually eclipses. The orbital inclination and fractional radii of the component 
stars can be determined by the analysis of photometric light curves. Therefore, eclipsing binaries are unique targets for 
determining the masses and radii from their combined light curves and radial velocities analyses. Nevertheless, absolute 
radii were measured only for a rather small number of early-type B-stars which are members of eclipsing 
binary systems \citep{Hil04,Tor10,Iba03,Iban13}. Thus, we started a systematic observing 
program devoted to the spectroscopic study of close eclipsing binary systems with at least one hot component. 

\subsection{V1441\,Aquilae} 
The eclipsing character of V1441\,Aquilae (HD 177624; BD+09$^{o}$3979; HIP 93732; V=6.90, B-V=0.19 mag) was 
discovered by the  $Hipparcos$ satellite mission \citep{Per97}, the primary eclipse having an amplitude of 
0.09\,mag. The depth of the secondary minimum is nearly half of the primary one. V1441\,Aql was first 
recognized as a double-lined spectroscopic binary by \citet{Hil80}. They classified it as a B3V star with 
semi-amplitudes of radial velocities of about 82 and 196 \kms. The first eclipse light curve (LC) was roughly 
revealed by the $Hipparcos$ observations. \citet{Kaz99} designated it as V1441\,Aql in the {\sc 74th Name-list 
of Variable Stars}, and classified it as an EB. In recent years, large-scale photometric surveys such as 
All Sky Automated Survey (ASAS, \citealt{Poj02}) have been conducted with the main aim of looking for transiting 
exoplanets. The valuable by-product of these searches has been the very large number of well sampled eclipsing 
binary LCs. One of the eclipsing binaries observed at the Las Campanas Observatory as part of the 
ASAS was V1441\,Aql. Although the phase coverage is very good the photometric accuracy of the observations 
is only around 0.05 mag.

The orbital period of V1441\,Aql was determined from the spectroscopic observations as 2.374148 days by \citet{Hil80}.
The first photometric observations of the system were made by the $Hipparcos$ mission and an orbital period of about
2.374 days was estimated \citep{Lef09}. The  $ASAS$ photometry permits a redetermination of the orbital period of the system.
A periodogram analysis performed with PERIOD04 \citep{Len05}, which was applied 
to all the data obtained by $ASAS$. We derive the following ephemeris 

\begin{equation}
Min I(HJD)=2\,455\,100.4153(32)+2^d.3741896(16) \times E
\end{equation}
where the standard deviations in the last significant digits are given in parentheses. 

\section{Observation}
Optical spectroscopic observations of V1441\,Aql was obtained at the TUBITAK National Observatory using
the Turkish Faint Object Spectrograph Camera (TFOSC)\footnote{http://http://tug.tubitak.gov.tr/tr/teleskoplar/tfosc}
attached to the 1.5\,m telescope. The observations were made from July 22, 2012 to August 3, 2013, under good seeing
conditions. Further details on the telescope and the spectrograph can be found at http://www.tug.tubitak.gov.tr. The 
wavelength coverage of each spectrum was 4000-9000 \AA~in 11 orders, with a resolving power of 
$\lambda$/$\Delta \lambda$ $\sim$7\,000 at 6500 \AA. The average signal-to-noise ratio (S/N) was $\sim$120. We also 
obtained high S/N spectra of early type standard stars 1\,Cas (B0.5\,IV), HR\,153 (B2\,IV), $\tau$\,Her (B5\,IV), 
21\,Peg (B9.5\,V) and $\alpha$\,Lyr (A0\,V) which were used as templates in derivation of the radial velocities. 

The echelle spectra were extracted from the raw images following standard reduction steps involving electronic 
bias subtraction, flat field division, cosmic rays removal, optimal extraction of the echelle orders, and 
wavelength calibration thanks to the emission lines of a Th-Ar lamp.  The reduction was performed using tasks 
of the IRAF package\footnote{IRAF is distributed by the National Optical Astronomy Observatory, which is
operated by the Association of Universities for Research in Astronomy, Inc. (AURA), under cooperative 
agreement with the National Science Foundation.}

\section{Radial velocities and atmospheric parameters}
The time-resolved spectroscopic dataset consists of 17 observations for V1441\,Aql. We have measured radial velocities 
(RVs) from the spectra, focusing on spectral segments containing the He\,{\sc i} $\lambda$5876 (order\,4) and $\lambda$6678 
(order\,3) lines which are the most prominent un-blended features in our spectra, apart from the Balmer lines. We have 
employed the standard cross-correlation method for measuring the velocities of the component stars of the 
systems. The numerical cross-correlation technique \citep{Sim74,Ton79} is a standard approach for measuring RVs 
from the spectra of close binary systems. Cross-correlation analyses were made using the spectra of $\tau$\,Her 
and 21\,Peg as templates. The principle spectral features showing splitting due to binarity were the  He\,{\sc i} 
lines at $\lambda\lambda$5876 and 6678. We used also order 9, containing the He\,{\sc i} $\lambda$4471 line, for 
a few measurements of the radial velocities. The spectra taken close to the conjunctions, which display no 
double-lined feature, were disregarded. The Balmer lines were not used in the measurements of radial velocities 
due to their extremely broad profiles. 

We obtained 17 radial velocities for each component of V1441\,Aql. The average radial velocities and their associated 
standard errors derived from the spectral segments containing He\,{\sc i} $\lambda\lambda$4471, 5876, and 6678 lines 
are presented in Table\,1, along with the observation date and orbital phase. The mean error of radial velocities is 
3.6\,km\,s$^{-1}$ for the primary, and 8.5 km\,s$^{-1}$ for the secondary star of V1441\,Aql. The RVs are plotted 
against the orbital phase in Fig.\,1, where the filled squares represent the primaries and the empty squares the 
secondaries, respectively. Examination of the $ASAS$ light curve show no evidence for any eccentricity in the orbits 
of both systems. Therefore, we have assumed circular orbits and analysed the RVs using the {\sc RVSIM} 
software programme \citep{Kan07}. Final orbital parameters are presented in Table\,2. 

Intermediate-resolution optical spectroscopy permits us to derive most of the fundamental stellar parameters, such as the
projected rotational velocity ($v\,sin\,i$), spectral type (S$_p$), luminosity class, effective temperature 
($T_{\rm eff}$), surface gravity ($log~g$), and metallicity ([Fe/H]).

The width of the cross-correlation function (CCF) is a good tool for the measurement of projected rotational velocity 
($v\sin i$) of a star. We use a method developed by \citet{Pen96} to estimate the $v\sin i$ of each star composing the 
investigated SB2 systems from its CCF peak by a proper calibration based on a spectrum of a narrow-lined star with a 
similar spectral type. Per each system, the rotational velocities of the components were obtained by measuring the 
FWHM of the CCF peak related to each component in five high-S/N spectra acquired near the quadratures, where the 
spectral lines have the largest Doppler-shift. The CCFs were used for the determination of $v\,sin\,i$ through a 
calibration of the full-width at half maximum (FWHM) of the CCF peak as a function of the $v\,sin\,i$ of artificially 
broadened spectra of slowly rotating standard star (21\,Peg, $v\sin i \simeq$14 km\,s$^{-1}$, e.g., \citealt{Roy02}) 
acquired with the same setup and in the same observing night as the targets systems. The limb darkening coefficient 
was fixed at the theoretically predicted values, 0.42 for both systems \citep{Van93}. We calibrated the relationship 
between the CCF Gaussian width and $v\,sin\,i$ using the \citet{Con77} data sample. This analysis yielded projected 
rotational velocities for the components of V1441\,Aql as $v_{\rm P}\sin i$=196~km\,s$^{-1}$, 
and $v_{\rm S}\sin i=101$~km\,s$^{-1}$. The mean deviations were 4 and 7 km\,s$^{-1}$, for the primary and 
secondary, respectively, between the measured velocities for different lines. 

\begin{figure}
\center
\hspace{-1.2cm}
\includegraphics[width=9.5cm,angle=0]{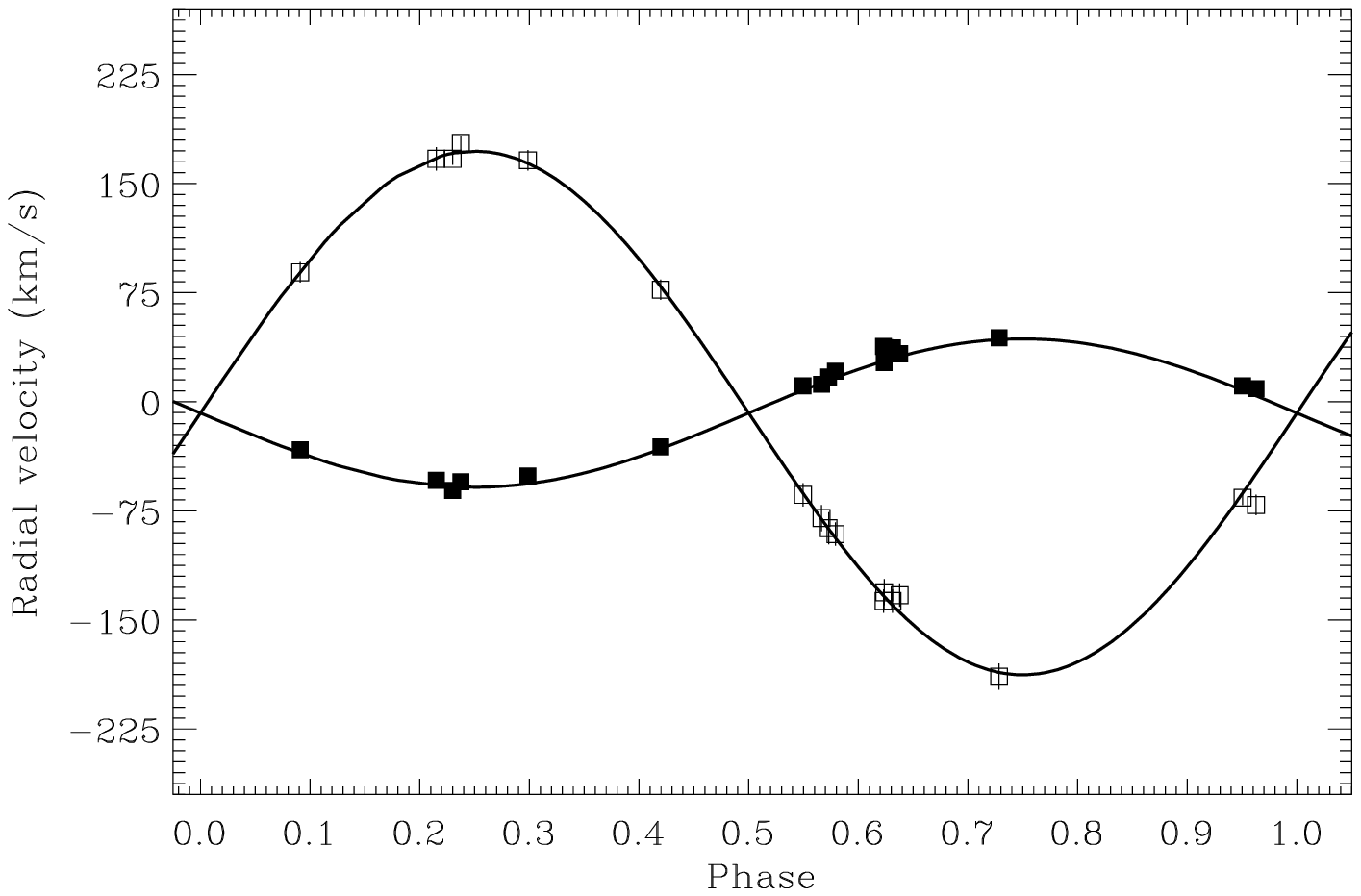}
\caption{Radial velocities for the components of V1441\,Aql. Filled squares correspond to the radial velocities 
for the primary and the empty squares for the secondary star. Error bars are shown by vertical line segments, 
which are smaller than symbol sizes. The solid lines are the computed radial velocity curves for the component stars.} 
\end{figure}

\begin{table}
\centering
\begin{minipage}{85mm}
\caption{Heliocentric radial velocities of V1441\,Aql. The columns give the heliocentric 
Julian date, orbital phase and the radial velocities of the two components with the corresponding 
standard deviations.}
\begin{tabular}{ccrcrcc}
\hline
HJD 2400000+ & Phase & \multicolumn{4}{c}{V1441\,Aql }&  	\\
             &       & $V_{\rm P}$                      & $\sigma$                    & $V_{\rm S}$   	& $\sigma$	\\
\hline
56131.3242	&	0.2137	&	$-54$	  &	  3	  &	  $ 167$    &	    8	    \\
56132.2925	&	0.6216	&	$ 38$	  &	  3	  &	  $-157$    &	    8	    \\
56133.4035	&	0.0895	&	$-33$	  &	  3	  &	  $  89$    &	    7	    \\
56134.4926	&	0.5483	&	$ 11$	  &	  2	  &	  $ -54$    &	    8	    \\
56135.4443	&	0.9491	&	$ 11$	  &	  2	  &	  $ -66$    &	    5	    \\
56136.2708	&	0.2972	&	$-51$	  &	  5	  &	  $ 166$    &	    7	    \\
56137.2913	&	0.7270	&	$ 44$	  &	  5	  &	  $-189$    &	    9	    \\
56147.3444	&	0.9614	&	$  9$	  &	  3	  &	  $ -71$    &	    7	    \\
56167.4232	&	0.4185	&	$-31$	  &	  2	  &	  $  77$    &	    7	    \\
56506.4817	&	0.2287	&	$-61$	  &	  3	  &	  $ 167$    &	    4	    \\
56506.4993	&	0.2361	&	$-55$	  &	  3	  &	  $ 178$    &	    5	    \\
56507.2803	&	0.5651	&	$ 12$	  &	  4	  &	  $ -60$    &	    9	    \\
56507.2957	&	0.5716	&	$ 17$	  &	  4	  &	  $ -67$    &	    11      \\
56507.3107	&	0.5779	&	$ 21$	  &	  4	  &	  $ -91$    &	    8	    \\
56507.4161	&	0.6222	&	$ 27$	  &	  4	  &	  $-141$    &	    9	    \\
56507.4338	&	0.6297	&	$ 37$	  &	  6	  &	  $-137$    &	    8	    \\
56507.4494	&	0.6363	&	$ 33$	  &	  5	  &	  $-133$    &	    8	    \\
\hline
\end{tabular}
\end{minipage}
\end{table}

\begin{table}
\centering
\begin{minipage}{85mm}
\caption {Results of the radial velocity analysis for V1441\,Aql.}
\begin{tabular}{@{}ccccccccc@{}c}
\hline
Parameter  & \multicolumn{2}{c}{V1441\,Aql }&  	\\
             & Primary           & Secondary   	\\
\hline
 $k$ (km\,s$^{-1}$) 							& 44$\pm$2 	     	&184$\pm$4		\\
    $V_\gamma$ (km\,s$^{-1}$) 						&\multicolumn{2}{c}{$-3.5\pm$0.2}  \\ 
   Average O-C (km\,s$^{-1}$)						& 3.6         	     	&7.5      	        \\    
   $a\sin i$ ($R_{\odot}$)						& 2.064$\pm$0.002	& 8.631$\pm$0.006     	\\ 
   $M\sin^3i$ ($M_{\odot}$)						& 2.353$\pm$0.140	& 0.563$\pm$0.041	\\  
\hline
\end{tabular}
\end{minipage}
\end{table}

We also performed a spectral classification for the components of the systems using COMPO2, an IDL code for the analysis 
of high-resolution spectra of SB2 systems written by one of us \citep[see, e.g.,][]{Frasca2006} and adapted to the TFOSC spectra. 
This code searches for the best combination of two reference spectra able to reproduce the observed spectrum of the system. 
We give, as input parameters, the radial velocities and projected rotational velocities $v\sin i$ of the two components 
of each system, which were already derived. The code then finds, for the selected spectral region, the spectral 
types and fractional flux contributions that better reproduce the observed spectrum, i.e. which minimize 
the residuals in the collection of difference (observed\,$-$\,composite) spectra. For this task we used reference 
spectra taken from the \citet{Val04} $Indo-U.S.\ Library\ of\ Coude\ Feed\ Stellar\ Spectra$ (with a a resolution of $\approx$\,1\AA)
that are representative of stars with spectral types from late-O type to early-A, and luminosity classes V, IV, and III. 
The atmospheric parameters of these reference stars were recently revised by \citet{Wu2011}.

We selected 198 reference spectra spanning the ranges of expected atmospheric parameters, which 
means that we have searched for the best combination of spectra among 39\,204 possibilities per each 
spectrum. The observed spectra of V1441\,Aql in the $\lambda\lambda$6525--6720 spectral region were best 
represented by the combination of HD\,179761 (B8 II-III) and HD\,182568 (B3 IV). However, we have adopted, 
for each component, the spectral type and luminosity class with the highest score in the collection of the 
best combinations of templates, where the score takes into account the goodness of the fit expressed by the 
minimum of the residuals. We have thus derived a spectral type of B3 subgiant for the primary and B9 giant 
for the secondary star of V1441\,Aql, with an uncertainty of about 1 spectral subclass. The effective temperature 
and surface gravity of the two components of each system are obtained as the weighted average of the values of the 
best spectra at phases near to the quadratures combinations of templates adopting a weight $w_i=1/\sigma_i^2$, where
$\sigma_i$ is the average of residuals for the $i$-th combination. The standard error of the weighted mean was 
adopted for the atmospheric parameters. Both stars appear to have a solar metallicity, within the errors. The 
atmospheric parameters obtained by the code and their standard errors are reported in Table\,3. The observed
spectra of V1441\,Aql at phases near to the quadratures are shown in Fig.\,2 together with the combination of two 
reference spectra which gives the best match.

\begin{table}
\centering
\begin{minipage}{85mm}
\caption {Spectral types, effective temperatures, surface gravities, and rotational velocities of the components 
estimated from the spectra of V1441\,Aql.}
\begin{tabular}{@{}ccccccccc@{}c}
\hline
Parameter  & \multicolumn{2}{c}{\sf V1441\,Aql }&  	\\
             & Primary           & Secondary   	\\
\hline
 Spectral type 			& B(3$\pm$0.5)\,IV 	     	&B(9$\pm$0.5)\,III	\\
 $T_{\rm eff}$ (K)	    	&18\,760$\pm$950  	     	&11\,670$\pm$650	\\   
 $\log~g$ ($cgs$)		& 3.77$\pm$0.05         	&3.95$\pm$0.17       	\\    
 $v\sin i$ (km\,s$^{-1}$)  	&196$\pm$4	  		& 101$\pm$7     	\\ 
\hline
\end{tabular}
\end{minipage}
\end{table}

\begin{figure}
\hspace{-0.8cm}
  \begin{center}
   \includegraphics[width=12.5cm]{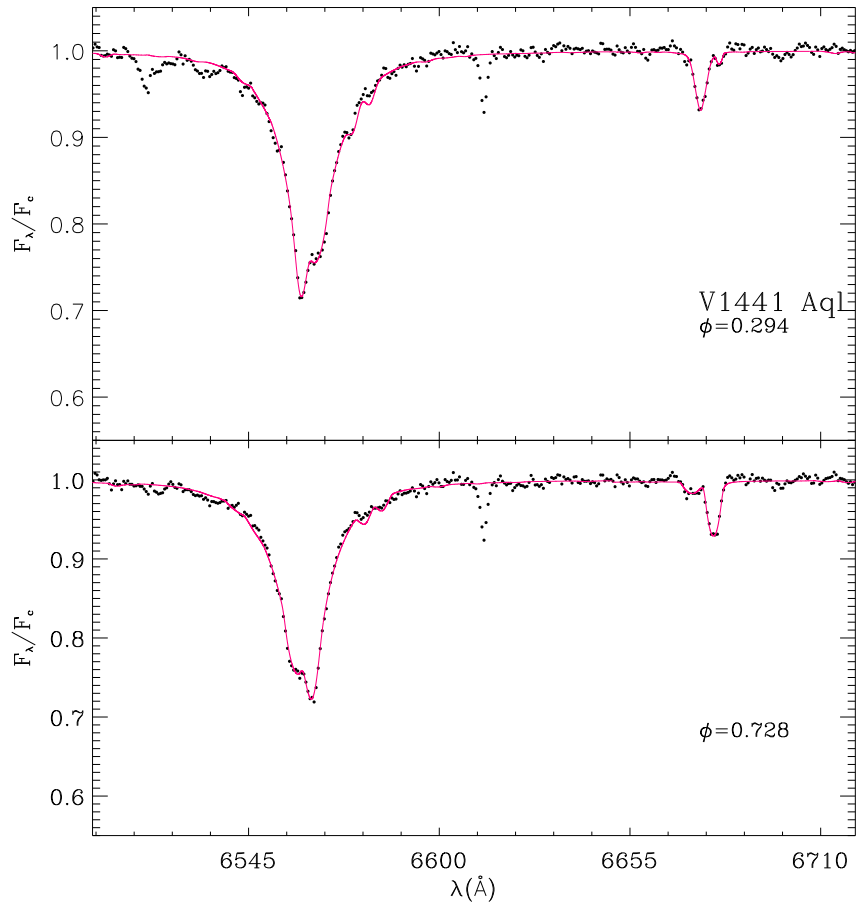}
  \end{center}
  \caption{Two spectra of V1441\,Aql near opposing quadrature phases, at phases 0.294 and 0.728. The 
wavelength limits are 6525-6715 \AA, which include the H$\alpha$ and He\,{\sc i} $\lambda$6678 lines. The 
deeper lines in each spectra refer to the primary star. Vertical axis is the normalized flux.} 
\end{figure}

\section{Analyses of the light curve}
We extracted the V-band light curve for V1441\,Aql (ASAS190509+0938.5) from the All Sky Automated Survey database 
\citep{Poj02}. The light curve composed of 375 photometric points, has a very good phase coverage and displays 
two equally spaced minima whose depths are about 0.09 and 0.04 mag. The observed light curve is reminiscent of an 
ellipsoidal variable with tidally distorted stars. V1441\,Aql was also observed by the $Hipparcos$ satellite 
\citep{Per97}. 69 Hp magnitudes were obtained for V1441\,Aql. The light curve is similar to that obtained by 
$ASAS$ but its shape is poorly defined due to small number of measurements. \citet{Mal06} estimate an Hp magnitude 
of 6.91 at the light maximum and a primary minimum with a depth of 0.09 mag in the $Hipparcos$ light curve. The light 
curve obtained by the $ASAS$ is shown in Fig.\,3, where the vertical axis is the normalized flux. 

The effective temperature of the primary star and the mass-ratio of the system are key parameters needed for the 
analysis of the ligth curve. The effective temperatures of both components have already been determined from the 
spectra as 18\,760$\pm$950 K and 11\,670$\pm$650\,K as well as a mass-ratio of 0.239$\pm$0.009 was derived from 
the radial velocity curve solution. Another approach to estimating the temperature of the primary star is to use 
the observed colours of the system. The apparent visual magnitude and colour indices were given as mean of nine 
measures by \citet{Mor71} as $V$=6$^m$.90, ($U-B$)=$-$0$^m$.36, ($B-V$)=0$^m$.19 and ($V-R$)=0$^m$.23 with an 
uncertainty of $\pm$0.01\,mag. The quantity $Q$=($U-B$) $-$ 0.72$(B-V)$ of the Johnson's UBV photometric system 
is independent of interstellar extinction. We compute the reddening-free index as $Q$=$-$0.497 for the 
system, which corresponds to a B4 star \citep{Hov04}, consistent with the spectral classification from the 
spectra. The effective temperature for the primary star, derived from the spectra, corresponds to an intrinsic 
colour of ($B-V$)=$-$0$^m$.22. A preliminary analysis of the light curve yields a light ratio of $l_{s}$/$l_{p}$=0.15 
for the V-passband. Using the intrisic colour of the primary star, the light ratio and observed colour of 0$^m$.19 
we compute an intrinsic composite colour for the system as ($B-V$)=$-$0$^m$.21. Then we estimate the reddening 
for the system as $E_{(B-V)}$=0$^m$.40.

\begin{table}
\scriptsize
\centering
\begin{minipage}{85mm}
\caption{Final solution parameters for the double-contact model of V1441\,Aql. }
\begin{tabular}{@{}ccccc}
\hline
Parameters  & V1441\,Aql \\                  
\hline
$i^{o}$			                &41.65$\pm$0.58		 \\
$T_{\rm eff_1}$ (K) 			&18\,760[Fix]		 \\
$T_{\rm eff_2}$ (K) 			&11\,650$\pm$300	 \\
$\Omega_1$ 	          		&2.327$\pm$0.030	 \\
$\Omega_2$				&2.327$\pm$0.030	 \\
$r_1$				        &0.4557$\pm$0.0065	 \\
$r_2$		           	        &0.2620$\pm$0.0033	 \\
$\frac{L_{1}}{(L_{1}+L_{2})}$   	&0.8721$\pm$0.0035	 \\
$\sum(O-C)^{2}$			        &0.0370              	 \\	   
$N$		           	        &375			 \\    
$\sigma$			        &0.0100		         \\				   
\hline  
\end{tabular}
\end{minipage}
\end{table}

Since the $Hipparcos$ data contain only 69 H$_{p}$ points with uncertainties larger than the $ASAS$ ones, we did not 
attempt to analyze them for determination of the orbital parameters. We started to analyze the $ASAS$ light curve 
using the Wilson-Devinney code \citep[hereafter WD; e.g.,][]{Wil71,Wil79,Wil06} as implemented in the software 
{\sc phoebe} \citep{Prs05}. The WD code is widely used for determination of the orbital parameters of the eclipsing 
binaries. To run the code we need  some initial elements. The logarithmic limb-darkening coefficients were 
interpolated in effective temperature and surface gravity from the coefficients tabulated by \citet{Van93}. The 
initial limb-darkening coefficients were taken as $x_1$=0.45, $y_1$=0.24, $x_2$=0.59, $y_2$=0.29 for the hotter 
and cooler star, respectively. They are updated at every iteration. The gravity-brightening coefficients 
$g_1$=$g_2$=1.0 and albedos $A_1$=$A_2$=1.0 were fixed for both components, as appropriate for stars with 
radiative atmospheres. We started with the $Mode-2$ option of the WD code, designed for detached binaries, for 
the analysis of light curves as described by \citet{Wil06}. The adjustable parameters in the light curve fitting 
were the orbital inclination $i$, the effective temperature of the secondary star $T_{\rm eff_2}$, the monochromatic 
luminosity of the primary L$_1$, and the zero-epoch offset. We could not obtain a good fit of the light curve with 
a detached configuration. This led us to try with a semi-detached configuration. We thus applied $Mode-5$ option 
of the WD code, in which the secondary star fills its lobe. The convergence was obtained with the primary star 
overfilling its lobe. Then we tried the $Mode-3$ (for overcontact systems, the stars are in geometrical contact 
without being in thermal contact). This solution resulted in that both stars are well inside their lobes. Then we 
applied $Mode-4$ (primary star fills its lobe) but we arrived at a result that the secondary star is overfilling 
its lobe. 

Finally, the solution with the $Mode-6$ option (for double contact systems, DCSs) resulted in acceptable parameters. 
\citet{Wil79} defined DCSs as a binaries in which both stars fill their Roche lobes and at least one rotates faster 
than synchronously, so that the components do not touch, even at one point. The fits of the computed light curves 
obtained with $Mode-6$ to the observations are satisfactory with the smallest sum of residuals squared. The out-of-eclipse 
part of the observed light curve is now better reproduced. We have already measured the projected rotational 
velocities from the spectral lines for the components of V1441\,Aql. Preliminary analysis using the $Mode-6$ 
of WD code yielded orbital parameters for the component stars. We have computed projected rotational velocities 
of the components. A comparison of the observed rotational velocities with those computed velocities shows that 
the stars rotate faster than the synchronous values. Therefore we repeated our analysis taking $F_1$=1.8 and $F_2$=1.6.

The parameters of our final solutions are listed in Table\,4. The orbital inclination of the system, effective 
temperatures, fractional mean radii (equivalent volume) of the components and fractional luminosity of the primary 
star in the V band were given in this table. The uncertainties assigned to the adjusted parameters are the 
internal errors provided directly by the code. The squared sum of residuals, $\sum(O-C)^{2}$, the number of 
data points, $N$, and the standard deviation, $\sigma$, of the observed light curve are quoted in the last three 
lines of Table\,4, respectively. The computed light curve is overplotted to the observations in Fig.\,3. This solution 
indicates for V1441\,Aql a grazing eclipse lasting about half an hour. 

\begin{figure}
  \begin{center}
   \includegraphics[width=12.5cm]{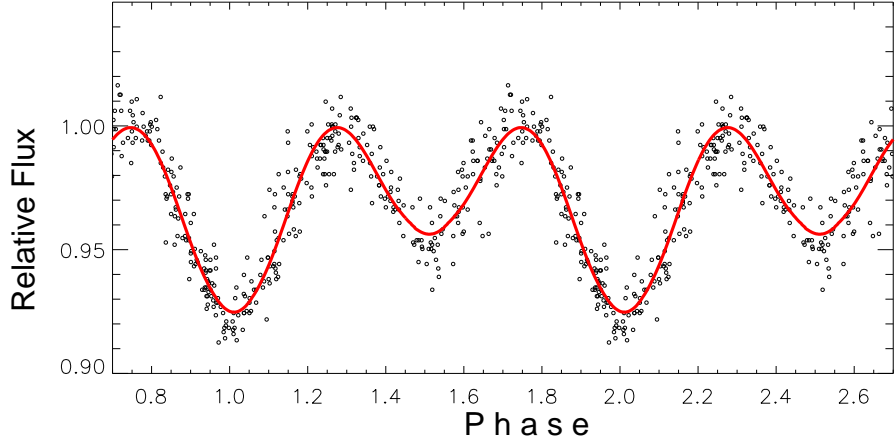}
  \end{center}
  \caption{Observed (dots) and computed ASAS V-band light curve of V1441\,Aql. The vertical axis is the normalized flux and 
the abscissa is the orbital phase. }
\end{figure}

\section{Results and discussion}   
Based on the results of radial velocities and light curves analyses we have calculated the physical properties 
of the V1441\,Aql. For this purpose, we used the $JKTABSDIM$ code developed by \citet{Sou05}. This code is now 
widely used for derivation of the absolute parameters of the eclipsing binary stars' components. It calculates 
complete error budgets using a perturbation algorithm. The fundamental stellar parameters for the components 
such as masses, radii, luminosities and their standard deviations have been derived using this code. The 
astrophysical parameters of the components, and other properties for the components of V1441\,Aql are presented
in Table\,5.

The separation between the components was found to be 16.09$\pm$0.34 R$_{\odot}$ for V1441\,Aql. The masses were 
measured to precision of about 6--7\,\% for the compoents. On the other hand the radii of the components have been 
derived with a precision of better than 5\,\%. The accuracy of any parameter of an eclipsing binary system depends 
mainly on the coverage of the both spectroscopic and photometric observations and their precision. In addition, the 
light curve solutions are more accurate for totally eclipsing systems. The light curve of V1441\,Aql shows instead 
a very shallow grazing eclipse. Despite these drawbacks, the physical parameters of the components of system could 
be determined with sufficient precision. We note that the effective temperatures of the secondary stars derived from 
the spectra are in good agreement with those obtained from the light curve analyses.

The luminosities and absolute bolometric magnitudes are calculated directly from the radii and and effective
temperatures of the components. The effective temperature of 5\,777~K and the absolute bolometric magnitude of
4.74 mag were adopted for the Sun \citep[e.g.,][]{Dri00}. The bolometric corrections were interpolated from the
tables of \citet{Flo96}. The V-band magnitudes of the systems at out-of-eclipse phases are taken as 6.90 and 7.32
for V1441\,Aql. We have calculated the absolute visual magnitudes for the components using the fractional 
luminosities and bolometric corrections given in Table\,4 and 5. Combining these values with the interstellar 
absorption of 1.27\,mag for V1441\,Aql and we have estimated the distances to the systems as 550$\pm$25.

In the log $T_{\rm eff}$--$\log L/L_{\odot}$ (left panel) and log $T_{\rm eff}$-log\,g planes (right panel) we 
have plotted the positions of the components (Fig.\,4), with 1-$\sigma$ error bars. The filled and empty squares 
refer to the components of V1441\,Aql. The evolutionary tracks and isochrones for the non-rotating single stars 
with solar composition are taken from \citet{Eks12}. 

The most striking result from analyses of the radial velocities and light curve is the double-contact configuration 
of V1441\,Aql. Double-contact systems (DCSs) were described in detail by \citet{Wil79}. He showed that in the 
double-contact systems the massive stars do not rotate synchronously. \citet{Wil85} estimated for the first time 
the rotation rate of a star in an eclipsing binary using photometric observations. They showed that the primary 
star of the double-contact system RZ\,Sct rotates about at 6.7 times the synchronous value. Very recently, 
\citet{Ter14} discussed the double-contact nature of the eclipsing binary system TT\,Her. They have already 
shown that the observed radial velocities and light curves could only be represented with a non-synchronous 
rotation of the primary star. Their analyses resulted in that the  more massive star rotates at 1.25 times the 
synchronous value. Our measured projected rotational velocities for V1441\,Aql show that both components rotate 
faster than the synchronous values, amounting to about 1.7$v_{syn}\sin i$.

The comparison with the evolution of single stars, shown in Fig.\,4, indicates that the age of V1441\,Aql is
longer than 30~Myr. Both stars have evolved away from the main sequence, i.e. exhausted the hydrogen in their cores.
The primary star is probably in the shell hydrogen burning phase. While the more massive star appears to be 
crossing the Hertzsprung gap in the HR diagram, the less massive star is close to the base of giant branch. In 
addition, the secondary component is seen as an over-luminous and hotter star with respect to its mass. This 
may indicate that the secondary star lost the outer envelope during the evolution in past. At present, it has 
a larger radius and therefore has a higher luminosity, similar to the donors in the semi-detached Algol-type 
binaries.        

We calculated the synchronization time for V1441\,Aql as about 2~Myr using the hydrodynamic damping formalism of
\citet{Tas97}. The Tassouls' synchronization time is about 15 times shorter than the estimated age of the binary.
Both observational \citep{Gla08} and theoretical \citep{Der10} studies showed that the mass-gainers of the
semi-detached close binaries rotate faster than synchronous. There is a consensus that the most plausible
reason for this asynchronous rotation of the primaries is the high values of the mass transfer rate. \citet{Der10} 
recently showed a relation between mass transfer rate and asynchronous rotation velocity of the gainers in 
the Algol-type systems. The double-contact nature of V1441\,Aql with asynchronous components indicates that 
it has passed through rapid phase of mass transfer.       

\setlength{\tabcolsep}{3pt}
\begin{table*}
\footnotesize
\centering
\begin{minipage}{85mm}
\caption{Absolute parameters, magnitudes and colours for the components of V1441\,Aql. }
\begin{tabular}{@{}llcccccccc@{}}
\hline
Parameter & Units & \multicolumn{2}{c}{V1441\,Aql }	\\
          &    & Primary                      & Secondary                     \\
\hline
Mass                & M$_{\odot}$   & 8.02$\pm$0.51 	& 1.92$\pm$0.14       \\ 
Radius              & R$_{\odot}$   & 7.33$\pm$0.19	& 4.22$\pm$0.11       \\
$T_{\rm eff}$       & K	            & 18\,760$\pm$950	& 11\,650$\pm$300     \\
$\log(L/L_{\odot})$ &	   	    & 3.779$\pm$0.091	& 2.471$\pm$0.047     \\ 
$\logg$             & $cgs$	    & 3.611$\pm$0.016	& 3.471$\pm$0.023     \\
$Sp.Type$           &		    & B3IV		& B9III 	      \\
$M_{bol}$           & mag           & $-4.70\pm$0.23	& $-1.43\pm$0.12      \\
$BC$                & mag	    & $-1.75$	        & $-0.61$	      \\
$M_{V}$             & mag           & $-2.95\pm$0.11	& $-0.82\pm$0.07      \\
$v\sin i_{\rm cal}$ & km\,s$^{-1}$  & 104$\pm$3         &  60$\pm$2  	      \\
$v\sin i_{\rm obs}$ & km\,s$^{-1}$  & 196$\pm$4 	& 101$\pm$7  	      \\
$d$                 & pc            & 560$\pm$32        & 548$\pm$22 	      \\
\hline
\end{tabular}
\end{minipage}
\end{table*}

\begin{figure*}
  \begin{center}
 \includegraphics[width=14cm]{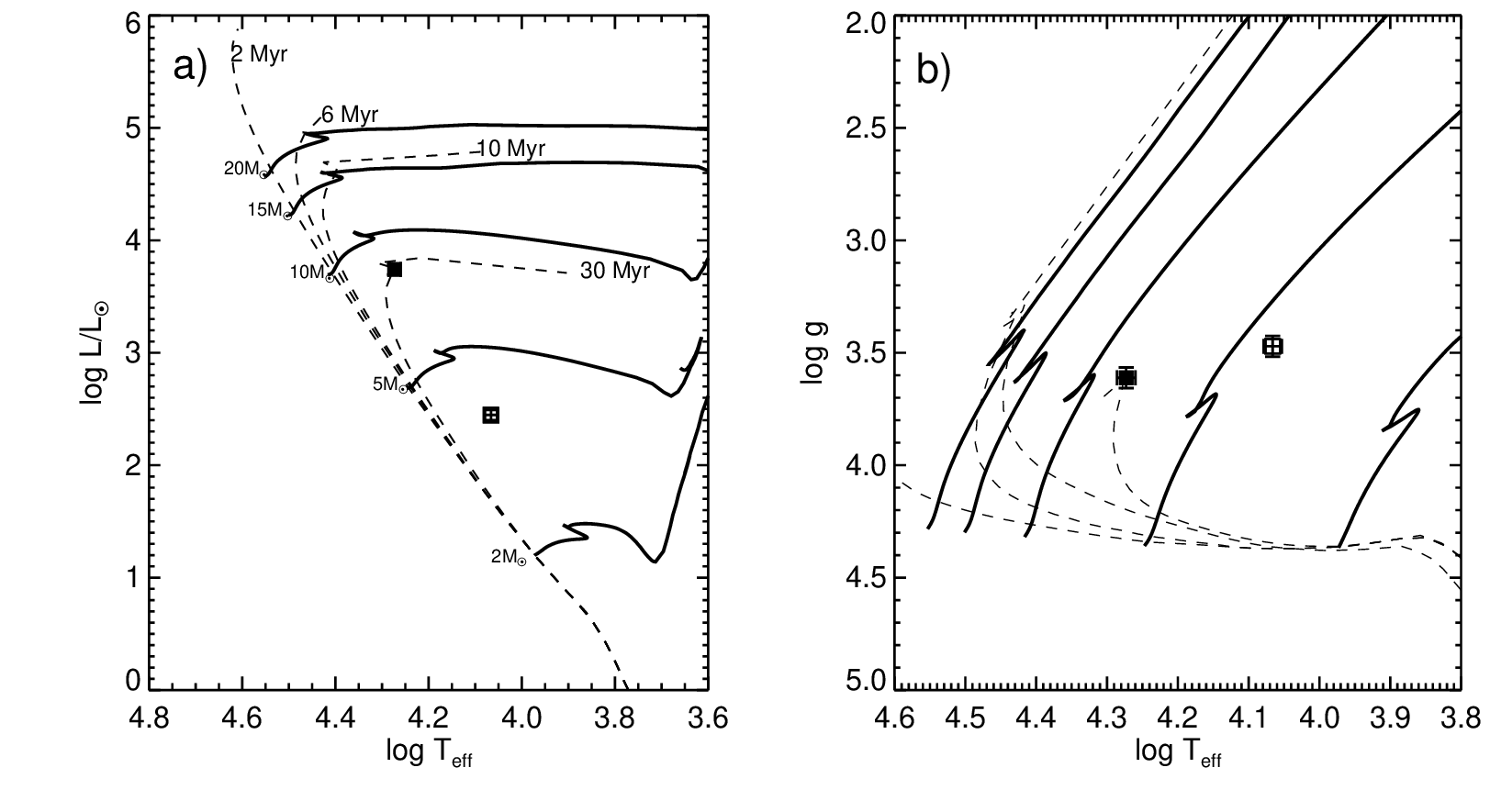}
  \end{center}
  \caption{Positions of the components of the systems in the luminosity-effective temperature and 
gravity effective temperature planes are plotted. The filled and empty squares refer the components of 
V1441\,Aql. The solid lines show evolutionary tracks for single stars with masses of 20, 15, 10, 5 
and 2 solar masses for solar composition taken from \citet{Eks12}. The stars of V1441\,Aql are departed 
from the main-sequence, locating at the Hertzsprung gap and close to the base of giant branch 
with an age of about 30\,Myr.}  \label{fig:evo}
\end{figure*}

\section{Summary} 
V1441\,Aql is close eclipsing binaries containing high-mass stars. We carried out spectroscopic observations
of system. The atmospheric parameters of the components in the eclipsing pairs have been determined from their 
spectra. The spectra were analyzed using cross-correlation for measuring the radial velocities of both components 
and with an ad-hoc code for deriving their atmospheric parameters. Moreover, the $ASAS$ and $HIPPARCOS$ light 
curves were modeled using the WD code. The physical parameters for the system is measured to accuracies of 
6-7\,\% in mass, and 5\,\% in radius. V1441\,Aql seems to a double-contact system with asynchronously-rotating 
components. The distances to the systems V1441\,Aql is estimated as 550$\pm$25\,pc. A comparison of physical 
parameters of the components with the theoretical models of single stellar evolution models has been made. An 
age of about 30 Myr is estimated for V1441\,Aql. 

\section*{Acknowledgments}
We thank to T\"{U}B{\.I}TAK National Observatory (TUG) for a partial support in using RTT150 
telescope with project number {\sf 11BRTT150-198}.
We thank to EB{\.I}LTEM Ege University Research Center for a partial support with project number {\sf 2013/BIL/018}.
We also thank to the staff of the Bak{\i}rl{\i}tepe observing station for their warm hospitality. This study is 
supported by Turkish Scientific and Technology Council under project number {\sf 112T263}.
This research was also partly supported by the Scientific Research Projects Coordination Unit of
Istanbul University. Project number 3685. We thank Canakkale Onsekiz Mart University Astrophysics
Research Center and Ulupinar Observatory together with Istanbul University Observatory
Research and Application Center for their support and allowing use of IST60 telescope.
This work was partially supported by the Italian {\em Ministero dell'Istruzione, Universit\`a e  Ricerca} (MIUR).
The following internet-based resources were used in research for this paper: the NASA Astrophysics Data System; the 
SIMBAD database operated at CDS, Strasbourg, France; and the ar$\chi$iv scientific paper preprint 
service operated by Cornell University. 

\end{document}